\documentclass{article}
\usepackage{amsmath}
\usepackage{graphicx}

\begin{document}

\author{Patricio S. Letelier\thanks{e-mail: letelier@ime.unicamp.br}%
\ \ \ and\ \ Samuel R. Oliveira\thanks{e-mail: samuel@ime.unicamp.br}\\
Departamento de Matem\'{a}tica Aplicada\\Universidade Estadual de Campinas\\13081-970 Campinas, S.P., Brazil.}
\title{On Uniformly Accelerated Black Holes}
\date{}
\maketitle
\begin{abstract}
The static and stationary C-metric are revisited in a generic framework and
their interpretations studied in some detail. Specially those with two event
horizons, one for the black hole and another for the acceleration. We found
that: i) The spacetime of an accelerated static black hole is plagued by
either conical singularities or lack of smoothness and compactness of the
black hole horizon; ii) By using standard black hole thermodynamics we show
that accelerated black holes have higher Hawking temperature than Unruh
temperature of the accelerated frame; iii) The usual upper bound on the
product of the mass and acceleration parameters ($<1/\sqrt{27}$) is just a
coordinate artifact. The main results are extended to accelerated rotating
black holes with no significant changes.
\end{abstract}

\section{Introduction}

Let us mention some relevant aspects of our present knowledge of black holes:
The uniqueness theorems\cite{Heusler96} lead us to the study of just two
families of exact solutions of Einstein equations for stationary vacuum
spacetimes-- the Schwarzschild's and Kerr's, and their charged versions.
Distortions and perturbations have been studied during the last two
decades\cite{GerochHartle82}\cite{Chandrasekhar83}. In the framework of
linearized approximations we learned that the holes response to external
perturbations appears as special modes of gravitational waves -- the quasi
normal ringing modes. Numerical simulations confirm that perturbed black holes
settle down by the emission of these modes\cite{NR}. There is strong evidence
for astrophysical black holes\cite{Rees98} which are perturbed by their environment.

There are also several open issues that have been presented as conjectures:
The cosmic censorship conjecture\cite{HawkingPenrose70}, the hoop conjecture
\cite{Thorne72}, the no-hair conjecture \cite{Israel87}, the topological
censorship conjecture \cite{Galloway93} and the adiabatic invariant conjecture
\cite{Bekenstein80}. Others have been studied in connection with
thermodynamics, statistical mechanics, quantum theory and cosmology
\cite{Wald98}. Also examples of more general black holes has been studied in
the context of supergravity, string theory and related theories
\cite{superhole}.

In this article we study some aspects of accelerated black holes. An
interesting feature of these holes is that from a semi-classical view point
both Hawking and Unruh radiation may be present because of the horizons
associated to the holes and to the acceleration.

The object of our study is an old exact solution of vacuum Einstein equations
found in 1917 by Levi-Civit\`{a} \cite{Levi-Civita18} and Weyl \cite{Weyl19}.
It is a simple and rich geometry. In the seventies, it was found a broader
class of exact solution with acceleration and rotation parameters
\cite{KinnersleyWalker70} \cite{PlebanskiDemianski76} and it was named as
C-metric or Weyl C-metric. The solutions were obtained by studying the
algebraic properties of special class of geometries. In the eighties, it was
known to belong to a general class of boost-rotation symmetric
spacetimes\cite{BicakSchmidt89}\cite{Bicak00}. These solutions have also
charged versions \cite{KinnersleyWalker70} \cite{FarhooshZimmerman79}. The
charged C-metric is interpreted as the solution for Einstein-Maxwell equations
for a charged particle moving with uniform acceleration
\cite{KinnersleyWalker70}. Another possible interpretation is the spacetime of
two Schwarzschild-type particles joined by a spring moving with uniform
acceleration \cite{Bonnor83}.

Actually, the C-metric can be associated to several spacetimes
\cite{FarhooshZimmerman80}. We review them, in section 2, using a slightly
different approach. The most interesting ones have two event horizons and a
point singularity. One event horizon has finite area, associated to a black
hole and the other event horizon has infinite area, associated to the Rindler
horizon of accelerated frames \cite{CornishUttley95}. We compute the surface
gravity on these horizons and conclude that, in general, the gravity at the
hole is larger than the frame acceleration. We show also, for generic
configurations, that the hole's horizons are not smooth compact surfaces and
confirm the well known fact that the line of acceleration is not elementary
flat. We remark that the product of surface gravity by the area of the horizon
gives exactly the expected mass of the hole. This result is expected because
of the coordinate transformation that map the C-metric into a Weyl solution
which is a superposition of a hole, with a given mass, and a semi-infinite rod
of linear density $1/2$ \cite{LetelierOliveira98}. Our units are such that
$c=G=1$. Finally we notice that the C-metric solution brings no limitation on
the acceleration of a black hole. The usual presentation of the solution has
the constraint $mA<1/\sqrt{27}$ where $m$ and $A$ are the mass and the
acceleration parameters. We show that this constraint is due sole to the
choice of coordinates.

In section 3 the rotating C-metric is studied in a similar way. The main new
features introduced by a rotation parameter is that it opens the possibility
of existence of ergoregions, spinning strings and spinning struts. We extend
most of the results of the previous section to include a rotation. The
interpretation of the more significant parts of the rotating C-metric is that
of a spacetime in the neighborhood of an accelerated Kerr-like particle
\cite{FarhooshZimmerman80b}. We show also that the internal singularity
resembles a rotating ring as in the standard Kerr solution.

The amount of gravitational radiation by accelerated black hole is not
computed in these paper \cite{Tomimatsu98}. As stationary solutions both C and
the rotating C metric represent eternal black holes being eternally
accelerated with gravitational wave coming from the singularities at infinity
in a such amount to balance the output of the accelerated black holes.

In the last section we summarize our main results and make some final comments.

\section{The C-metric}

Let us first review, in a more general framework, the physical meaning of the
vacuum C-metric \cite{Levi-Civita18} \cite{NewmanTamburino61}
\cite{FarhooshZimmerman80} whose line element is
\begin{equation}
ds^{2}=\frac{1}{A^{2}\left(  x+y\right)  ^{2}}\left[  K^{2}F(y)dt^{2}%
-\frac{dy^{2}}{F(y)}-\frac{dx^{2}}{G(x)}-\frac{G(x)}{K^{2}}d\phi^{2}\right]  .
\label{cmetric}%
\end{equation}
All the coordinates and the constant $K$ are dimensionless. The constant $A$
has dimension of inverse of length, which is used to fix the scale of physical
interest. The functions $G(x)$ and $F(y)$ are cubic functions such that
$G(x)=-F(-x)$. Let us consider the real cubic $Q$ of a real variable $w$%
\begin{equation}
Q(w)=\alpha\left(  w-w_{1}\right)  \left(  w-w_{2}\right)  \left(
w-w_{3}\right)  . \label{Q3}%
\end{equation}
Let us assume $\alpha>0$ and $Q(w)$ has three real roots $w_{1}<w_{2}<w_{3}$.
Setting $G(x)=Q(-x)$ and $F(y)=-Q(y)$, the infinity $x-y$ plane is divided
into 16 rectangular regions. Let us suppose the $x$'s range is such that
$G(x)\geq0$. Then, for $-\infty<y<+\infty$ the metric function $F(y)$ changes
sign on the roots $w_{1}$, $w_{2}$ and $w_{3}$ and the type of the coordinates
$t$ and $y$ are interchanged between time-like and space-like. Now, let us
suppose the $y$'s range is such that $F(y)\geq0$. Then, as $-\infty<x<+\infty$
the other metric function $G(x)$ changes sign on the roots $-w_{3}<-w_{2}<-w_{1}$ 
and the signature of the metric (\ref{cmetric}) changes
between $-2$ and $+2$. The 2-dimensional spaces $t=const$, $y=w_{k}$, $k=1..3$
can have finite or infinite area which we compute below, while the
2-dimensional spaces $t=const$, $x=-w_{k}$, $k=1..3$ has a vanishing area,
that is, it is degenerate into a line (or a point). We can estimate whether or
not the length of these lines are finite without knowing the roots explicitly.

\begin{table}[tbp] \centering

\begin{tabular}
[c]{|c|c|c|c|c|c|c|c|}\hline
$y\;\backslash x$ & $x<-w_{3}$ & -$w_{3}$ & (-$w_{3},$-$w_{2}$) & -$w_{2}$ &
(-$w_{2},$-$w_{1}$) & -$w_{1}$ & -$w_{1}<x$\\\hline
$y>w_{3}$ & -+--(1) & $\infty$ & -+++ & $\mathcal{L}$ & -+--(5) &
$\mathcal{L}$ & -+++\\\hline
$w_{3}$ & $\infty$ &  & $\infty$ &  & $\mathcal{A}$ &  & $\mathcal{A}$\\\hline
$(w_{2},w_{3})$ & +---(2) & $\infty$ & +-++ & $\infty$ & +---(6) &
$\mathcal{L}$ & +-++\\\hline
$w_{2}$ & $\mathcal{A}$ &  & $\infty$ &  & $\infty$ &  & $\mathcal{A}$\\\hline
$(w_{1},w_{2})$ & -+--(3) & $\mathcal{L}$ & -+++ & $\infty$ & -+--(7) &
$\infty$ & -+++\\\hline
$w_{1}$ & $\mathcal{A}$ &  & $\mathcal{A}$ &  & $\infty$ &  & $\infty$\\\hline
$w_{1}>y$ & +---(4) & $\mathcal{L}$ & +-++ & $\mathcal{L}$ & +---(8) &
$\infty$ & +-++\\\hline
\end{tabular}
\caption{
In the first column and in the first row the  $y$ and $x$ range are displayed.
The second, fourth, sixth and eight columns have the signature of the C-metric
depending on the sign of the functions $F(y)$ and $G(x)$.
On the roots $y=w_k$
the area of the event horizons are shown as finite $\mathcal{A}$ or $\infty$.
On the roots
$x=-w_k$ the lenght of the lines are shown as finite $\mathcal{L}%
$ or $\infty$.
The table is symmetric
about $x+y=0$. For comparison see similar tables in \protect\cite{CornishUttley95}
}
\end{table}

In Table I we present the signature associated to the metric (\ref{cmetric})
depending on the range of the coordinates $(t,y,x,\phi)$. The event horizons
associated to the Killing vector $\xi=A\partial_{t}$ are the roots of $F(y)$.
The regions on the same column are divided by Killing horizons at the roots
$y=w_{j},\;j=1,2,3$ . The regions on the same rows are disconnected because
they have different global signature. They are separated by the roots
$x=-w_{k},\;k=1,2,3$. We assume the range of the other coordinates as
$0\leq\phi\leq2\pi$ and $-\infty<t<\infty$ . Thus, the physically meaningful
spacetimes are those in which $\phi$ is a space-like coordinate. Thus the $x$
range has to be either $-\infty<x<-w_{3} $ or $-w_{2}<x<-w_{1}$ and the
associated spacetimes have signature $-2$. Therefore the regions where Killing
vector $\xi=A\partial_{t}$ is time-like represent static and axially symmetric
spacetimes and they must belong to the Weyl class\cite{Krameretal80}. One can
divide the $x-y$ plane in a similar way for the case $\alpha$ $<0$.

See also Figure 1.

The physical contents can be shown by the scalar invariants
\cite{CarminatiMcLenaghan91}. The simplest non-vanishing ones for the C-metric
are \cite{grtensor}
\begin{align}
C_{abcd}C^{abcd}  &  =12\alpha^{2}A^{4}\left(  x+y\right)  ^{6}\label{C2a}\\
C_{abcd}C^{cdef}C_{ef}^{\;\;ab}  &  =12\alpha^{3}A^{6}\left(  x+y\right)  ^{9}
\label{CI2a}%
\end{align}
where $C_{abcd}$ is the Weyl conformal tensor. Therefore, locally, the only
physically meaningful constants are $\alpha$ and $A$. They are called
dynamical parameters \cite{PlebanskiDemianski76} in contrast to the
kinematical ones: $w_{1},w_{2},w_{3}$ and $K$. Furthermore, the spacetimes are
not singular at the horizons. They have only singularities at $\left(
x+y\right)  \rightarrow\pm\infty$. We use below the notation $w_{0}=-\infty$
and $w_{4}=+\infty$.

We can introduce, for future convenience, another constant $m$ with dimension
of length such that
\[
\alpha=2mA.
\]
Therefore the spacetimes have two dimensional dynamical parameters $m$ and $A
$. They are associated to mass and acceleration parameters respectively. The
two independent limiting cases $m\rightarrow0$ and $A\rightarrow0$ have been
reported in the literature. The former is an accelerated frame while the
latter is a black hole. The cubic degenerates into a quadratic or a linear
function. A new justification of this interpretation is given below.

Let us compute the area of the horizons at $y=w_{j}$ where $F(y)=0$ by
integrating $x$ and $\phi$ in the their ranges
\begin{equation}
\mathcal{A}_{(j)}^{[k+1,k]}=\frac{2\pi}{A^{2}K}\int_{-w_{k+1}}^{-w_{k}}%
\frac{dx}{\left(  x+w_{j}\right)  ^{2}}=\frac{2\pi}{A^{2}K}\frac{w_{k+1}%
-w_{k}}{\left(  w_{j}-w_{k}\right)  \left(  w_{j}-w_{k+1}\right)  }
\label{area}%
\end{equation}
Some of the horizons have finite area ($j\neq k$ and $j\neq k+1$) so they are
black hole event horizons, while the infinity area ones are acceleration event
horizons. The area of the surfaces $y\rightarrow\pm\infty$ vanishes. The
symbolic values of the areas are indicated in the Table I and Figure 1.

One can also compute the distance between the horizons along the axis
$x=-w_{k}$ such that $dx=dt=0$ and $G(-w_{k})=0$.
\begin{equation}
\mathcal{L}_{(k)}^{[j+1,j]}=\frac{1}{A}\int_{w_{j}}^{w_{j+1}}\frac{dy}{\left|
y-w_{k}\right|  \sqrt{\left|  F(y)\right|  }} \label{lenght}%
\end{equation}
The possible values of the distances are presented in the Table I. They may
vanish, be infinite or have a finite value, say $\mathcal{L}$, according to
the convergence behavior of the integral in (\ref{lenght}).

The qualitative interpretation of the regions labeled by $1$ to $8$ in the
Table I is as follows. The regions $1-5$ and $8$ are spacetimes with essential
singularities. The odd labeled regions are not static. Note the region $3$: It
is a compact spacetime with two black holes separated by a finite distance on
one side and both boles attached to a singularity on the other side. Note the
regions 5 and 6: They represent the interior of a distorted black hole and the
exterior of an accelerated black hole respectively. The finite piece of the
axis is behind the black hole. In the literature there are some explicit
coordinate transformations from some patches of the C-metric to accelerated
black holes, double black holes at infinity, infinity black hole plus black
holes and so on \cite{Bonnor83} \cite{CornishUttley95}.

Let us suppose $Q(w)$ has only one real root $w^{\ast}$. As above, set
$G(x)=Q(-x)$ and $F(y)=-Q(y)$. Then the $x-y$ plane is divided into 4
rectangular regions.%

\begin{table}[tbp] \centering
\begin{tabular}
[c]{|c|c|c|c|}\hline
& $x<-w^{\ast}$ &  $x=-w^{\ast}$ & $x>-w^{\ast}$\\\hline
$y>w^{\ast}$ & $-+--$ & $\infty$ & $-+++$\\\hline
$y=w^{\ast}$ & $\infty$ &  & $\infty$\\\hline
$y<w^{\ast}$ & $+---$ & $\infty$ & $+-++$\\\hline
\end{tabular}
\caption{
In the first column and in the first row the  $y$ and $x$ range are displayed.
Single root case.
The second and fourth columns have the signature of the C-metric
for the
coordinates $(t,y,x,\phi)$
depending on the sign of the metric functions $F(y)$ and $G(x)$. On the root
$y=w^*$
the area of the event horizons are $\infty$. On the root
$x=-w^*$ the lenght of the lines are  $\infty$.
}
\end{table}

There is an infinite area horizon at $y=w^{\ast}$. The distances along
$x=-w^{\ast}$ are infinite. Assuming the same character for the coordinates
$t$ and $\phi$ as above, we restrict the meaningful spacetime to $x<-w^{\ast}%
$. The interpretation is that of an accelerated frame with conical
singularities along the line of acceleration and essential singularities at
infinity. See Table II.

There are of course other intermediate cases for the roots of $Q(w)$, but we
resume our discussion about the three real roots case.

We can compute the surface gravity $\kappa$ on the Killing horizons where the
Killing vector $\xi=A\partial_{t}$ vanishes, i.e.
\begin{align}
\kappa^{2}  &  \equiv\left.  -\frac{1}{2}\nabla_{\mu}\xi_{\beta}\nabla^{\mu
}\xi^{\beta}\right|  _{\left|  \xi\right|  =0}\label{surfgrav}\\
\kappa_{(i)}  &  =\frac{KA}{2}\left|  \frac{dF}{dy}\right|  _{y=w_{i}}\;
\label{surfgravb}%
\end{align}
Thus the dynamical parameter $A$ is proportional to the acceleration surface
gravity. Note that $\kappa_{1}>\kappa_{2}$ and $\kappa_{3}>\kappa_{2}$, that
is, the horizons at $y=w_{1}$ and $y=w_{3}$, which are ``closer to the
singularities'' at $y\rightarrow\pm\infty$, have stronger surface gravity than
the ``inner'' horizon at $y=w_{2}$. In particular for the region 6, the
surface gravity $\kappa_{(3)}$ at the black hole is larger than the
acceleration $\kappa_{(2)}$. Thus, using the semi-classical analogy between
$\kappa_{(3)}$ and the Hawking temperature of a black hole and between
$\kappa_{(2)}$ and the Unruh temperature of the accelerating frame one
concludes that the black hole is not in thermodynamical equilibrium with the
Unruh environment because of its higher temperature.

For generic black holes, the product of the surface gravity by the area of the
horizon is proportional to the mass of the hole\cite{GerochHartle82}. From
(\ref{surfgrav}) and (\ref{area}) we get
\begin{equation}
\kappa_{(i)}\mathcal{A}_{(i)}^{[k+1,k]}=4\pi\frac{m\left(  w_{k+1}%
-w_{k}\right)  }{2}=4\pi\;Mass \label{mass}%
\end{equation}
Thus the parameter $m$ is proportional to the mass of the hole.

The Killing axisymmetric vector $\eta=\partial_{\phi}$ has zero norm on the
axis of the symmetry.
\begin{equation}
\eta^{2}=\frac{G(x)}{\left[  KA\left(  x+y\right)  \right]  ^{2}}%
\end{equation}
Therefore, the roots of the cubic $G(x)$ are indeed the symmetry axis.

Based sole on the identification of the roots and $G(x)$ as the axis one can
compute the ratio between the length of a circle by $2\pi$ times its radius of
the metric (\ref{cmetric}). If this ratio is not unity, there is an angle
depletion, that is, a conical singularity.
\begin{equation}
\lim_{\varepsilon\rightarrow0}\frac{\int_{0}^{2\pi}\frac{1}{A\left|
-w_{i}+\varepsilon+y\right|  }\frac{\sqrt{\left|  G(-w_{i}+\varepsilon
)\right|  }}{K}d\phi}{2\pi\int_{-w_{i}}^{-w_{i}+\varepsilon}\frac{1}{A\left|
x+y\right|  }\frac{dx}{\sqrt{\left|  G(x)\right|  }}}=\frac{G_{x}\left(
-w_{i}\right)  }{2K} \label{con1}%
\end{equation}
One can choose the constant $K$ in such a way to avoid the conical singularity
in a particular piece of the axis. But in general the conical singularity will
show up somewhere on the axis. This is a known feature of the boost-rotation
symmetric spacetimes in which the C-metric is just one example\cite{Bicak00}.

It is also instructive to compute the Gaussian curvature $GC$ of the constant
$t$ and constant $y$ surface. It is given by%

\[
GC=A^{2}(2mA(x+y)^{3}+F(y))
\]
from which we can use the Gauss-Bonet theorem \cite{Manfredo} to obtain the
Euler characteristic $\chi$ of the horizon for $-w_{j}<x<-w_{j-1}$ at
$y=w_{i}$ where $F(y)=0$.
\begin{align}
\chi_{_{i}}^{[j,j-1]}+b.t.  &  =\frac{1}{2\pi}\int\int GC\frac{dxd\phi}%
{KA^{2}\left(  x+w_{i}\right)  ^{2}}\label{Gauss-Bonet}\\
&  =\frac{mA}{K}\left(  w_{j}-w_{j-1}\right)  \left[  2w_{i}-\left(
w_{j}+w_{j-1}\right)  \right] \nonumber
\end{align}
The boundary terms $b.t.$ vanish if the surface is a compact closed smooth
surface (CCSS) and the right-hand side of the equation above is an integer
number. It is clear that, in general, the horizons are not CCSS, unless we
adjust $K$ for this purpose. Of course we can only apply equation
(\ref{Gauss-Bonet}) if the surface is finite. Simple torus ($\chi=0$) black
holes are selected by choosing the roots such that $w_{j}=w_{j-1}$ or
$2w_{i}=w_{j}+w_{j+1}$, for example.

Thus the kinematical parameter $K$ can be chosen to either get rid of the
conical singularity in a piece the axis or to make the horizon a CCSS, but not
both. Using the membrane paradigm for the black holes and the vision of
conical singularities as struts or strings we conclude that in order to
accelerate a black hole one needs to push it with a strut and pull it with a
string carefully enough in order to not make a hole on its horizon. If one
just pushes or pulls it, the membrane will be somehow teared and the horizon
will not be a CCSS.

Let us focus on the region 6 of Table I: $w_{2}<y<w_{3}$ and $-w_{2}<x<-w_{1}
$. It is an accelerated frame with black hole. The Newtonian mass of the
finite line source with mass density $\frac{1}{2}$ is $m\left(  w_{2}%
-w_{1}\right)  /2$ which is exactly the mass of the hole (\ref{mass}) as
calculated above. The ratio between the surface gravity at $y=w_{3}$ to the
acceleration at $y=w_{2}$ is $\kappa_{(3)}/\kappa_{(2)}=\left(  w_{3}%
-w_{1}\right)  /\left(  w_{2}-w_{1}\right)  >1$, so the hole would evaporate
through the Hawking radiation despite the presence of the Unruh radiation of
the accelerated frame.

We can adjust the constant $K$ in three ways:

\begin{enumerate}
\item \textbf{Strut case}: There is a conical singularity at $x=-w_{1}$. Thus,
from (\ref{con1}) at $x=-w_{2}$ we get $K=mA\left(  w_{2}-w_{1}\right)
\left(  w_{3}-w_{2}\right)  $. The compression force (\ref{force}) on the
strut is $F_{z}=\frac{1}{4}\left(  w_{2}-w_{1}\right)  /\left(  w_{3}%
-w_{2}\right)  $. The Gauss-Bonet term (\ref{Gauss-Bonet}) at $y=w_{3}$
becomes $\left[  2w_{3}-\left(  w_{2}+w_{1}\right)  \right]  /\left(
w_{3}-w_{2}\right)  $ ; it is not an integer, in general. The area of the
finite horizon (\ref{area}) at $y=w_{3}$ is $\pi/\left(  A^{3}m\left(
w_{3}-w_{2}\right)  ^{2}\left(  w_{3}-w_{1}\right)  \right)  $. The surface
gravity (\ref{surfgrav}) at $y=w_{3}$ is $\kappa_{(3)}=mA^{2}\left(
w_{2}-w_{1}\right)  \left(  w_{3}-w_{2}\right)  ^{2}\left(  w_{3}%
-w_{1}\right)  $ .

\item \textbf{String case}: There is a conical singularity at $x=-w_{2}$.
Thus, from (\ref{con1}) at $x=-w_{1}$ we get $K=mA\left(  w_{3}-w_{1}\right)
\left(  w_{2}-w_{1}\right)  $. The compression force (\ref{force}) on the
string is $F_{z}=\frac{1}{4}\left(  w_{2}-w_{1}\right)  /\left(  w_{3}%
-w_{1}\right)  $. The Gauss-Bonet term (\ref{Gauss-Bonet}) at $y=w_{3}$
becomes $\left[  2w_{3}-\left(  w_{2}+w_{1}\right)  \right]  /\left(
w_{3}-w_{1}\right)  $; it is not an integer, in general. The area of the
finite horizon (\ref{area}) at $y=w_{3}$ is $\pi/\left[  A^{3}m\left(
w_{3}-w_{2}\right)  \left(  w_{3}-w_{1}\right)  ^{2}\right]  $. The surface
gravity (\ref{surfgrav}) at $y=w_{3}$ is $\kappa_{(3)}=mA^{2}\left(
w_{2}-w_{1}\right)  \left(  w_{3}-w_{2}\right)  \left(  w_{3}-w_{1}\right)
^{2}$.

\item \textbf{Smooth surface case}: From (\ref{Gauss-Bonet}) at $y=w_{3}$ we
fix
\begin{equation}
K=mA\left(  w_{2}-w_{1}\right)  \left[  2w_{3}-\left(  w_{2}+w_{1}\right)
\right]  /\chi
\end{equation}
for some integer number $\chi$ which is the Euler characteristic of the
horizon. There will be conical singularity at both $x=-w_{1}$ (strut) and
$x=-w_{2}$ (string). We can compute the compression force on them so that the
difference is given by $F_{1}-F_{2}=\frac{1}{4}\chi\left(  w_{2}-w_{1}\right)
/\left[  2w_{3}-\left(  w_{2}+w_{1}\right)  \right]  $. Both surface gravity
(\ref{surfgrav}) and the area of the horizon (\ref{area}) can be computed also.
\end{enumerate}

Thus the precise interpretation of this particular patch of the C metric could
be either that of i) an eternally accelerated eternal black hole with conical
singularities on the axis ahead \textit{and} behind the hole or ii) an
eternally accelerated eternal black hole with non smooth horizon with conical
singularities on the axis ahead \textit{or} behind the hole. The distortion of
the horizon due to the acceleration of the inertial frame has been
investigated \cite{FarhooshZimmerman80}.

The case of a double root at $y=w_{1}=w_{2}$ and another root at $y=w_{3}$
corresponds to an accelerated Chazy-Curzon particle\cite{Curzon24}%
\cite{BonnorSw64}\cite{Bicak00}. It is known that the Chazy-Curzon solution by
itself has directional singularity. The same is true for the accelerated case.
The other double root case: $y=w_{3}=w_{2}$ and another root at $y=w_{1}$
would correspond to the case when a black hole event horizon touches the
Rindler horizon \cite{Zhengetal97}. From the point of view of the geometry,
the limit $w_{3}\rightarrow w_{2}$ would lead to the equality of the surface
gravity at Schwarzschild and Rindler horizons meaning a thermodynamical
equilibrium of hole in the non-inertial frame \cite{Yi95}. The case of complex
conjugated roots and another real root would correspond to accelerated
Morgan-Morgan disk. All these cases are beyond the scope of this paper.

As presented here, there is no limitation on the values of $mA$ because we can
freely set the roots $w_{1},w_{2}$ and $w_{3}$. On the other hand if we set
the cubic to be $Q(w)=1-w^{2}+2mAw^{3}$, as usual in the literature, we need
the constraint $mA<1/\sqrt{27}$ to have three real roots otherwise the
solution will be that of an accelerated frame with no black holes. Then, $m$
and $A$ have no meaning by themselves.

See the Appendix for the connection between the C metric and the Weyl
coordinates for vacuum static axisymmetric spacetimes.

\section{Rotating C-metric}

Let us now present the metric that describes a spacetime of a uniformly
accelerating and rotating black hole in the same approach used above. It is
called the rotating vacuum C-metric\cite{FarhooshZimmerman80b}\cite{Bicak99}.

We expect three dimensional constants associated to the acceleration $A$, the
mass $m$ and the spin $a$ of the black hole. One version of this metric is
given by \cite{PlebanskiDemianski76}%

\begin{equation}%
\begin{array}
[c]{ll}%
ds^{2}=\frac{1}{A^{2}\left(  x+y\right)  ^{2}} & \left[  \frac{F(y)}{W}\left(
Kdt-\frac{aA}{K}x^{2}d\phi\right)  ^{2}-\frac{W}{F(y)}dy^{2}-\frac{W}%
{G(x)}dx^{2}\right. \\
& \left.  -\frac{G(x)}{W}\left(  \frac{1}{K}d\phi+aAy^{2}Kdt\right)
^{2}\right]
\end{array}
\label{cmetricstat}%
\end{equation}
\newline All the coordinates and the constant $K$ are dimensionless. The
constant $a$ has the dimension of length and $A$ of the inverse of length. The
functions $G(x)$ and $F(y)$ are quartic polynomials such that $G(x)=-F(-x)$
and
\begin{equation}
W\equiv1+\left(  aAxy\right)  ^{2}. \label{W}%
\end{equation}
Let us consider the real quartic $Q$ of a real variable $w$%
\begin{align}
Q(w)  &  =\delta+2An\;w+\varepsilon\;w^{2}+2Am\;w^{3}-\left(  aA\right)
^{2}\delta\;w^{4}\label{Q4}\\
&  =\alpha\left(  w-w_{1}\right)  \left(  w-w_{2}\right)  \left(
w-w_{3}\right)  \left(  w-w_{4}^{\ast}\right) \\
&  =\alpha\left(  w-w_{1}\right)  \left(  w-w_{2}\right)  \left(
w-w_{3}\right)  \left[  1+\left(  aAw_{2}\right)  ^{2}w_{3}w\right]
\end{align}
The roots $w_{1}$, $w_{2}$ and $w_{3}$ will be the relevant ones. We set below
$w_{1}=-w_{2}$ to simplify the expressions. The fourth root $w_{4}^{\ast}$ is
fixed by the others. The metric (\ref{cmetricstat}) becomes a vacuum solution
of Einstein equations by setting $G(x)=Q(-x)$ and $F(y)=-Q(y)$. Note that
$x=y=0$ have been picked up as a special point in this setup. The constants
$\delta$ and $\varepsilon$ are kinematical parameters while $a,A,m$ and $n$
are dynamical parameters as can be seen from the following invariants
\cite{grtensor} ($\beta\equiv aAn/m$ ).%

\begin{equation}%
\begin{array}
[c]{ll}%
C_{abcd}C^{abcd}=\,48m^{2}A^{6}\left(  \frac{x+y}{W}\right)  ^{6} & \left[
\left(  1-\beta^{2}\right)  \left(  W^{2}-16W+16\right)  \right. \\
& \left.  +4\beta\left(  3W-4\right)  \left(  W-4\right)  \right]
\end{array}
\label{C2b}%
\end{equation}
and the product the Weyl tensor with its dual%

\begin{equation}%
\begin{array}
[c]{ll}%
C_{abcd}^{\ast}C^{abcd}=96m^{2}A^{6}\,\left(  \frac{x+y}{W}\right)  ^{6} &
\left[  \left(  1-\beta^{2}\right)  aAxy\left(  3W-4\right)  \left(
W-4\right)  \right. \\
& \left.  +\beta\left(  8W^{2}-19W+12\right)  \right]
\end{array}
\label{CI2b}%
\end{equation}
Compare (\ref{C2b}) with (\ref{C2a}). The\ singularities appear only at
$\left(  x+y\right)  /W\rightarrow\pm\infty$ . If $a\neq0$ these singularities
are the points $(0,\pm\infty)$ and $(\pm\infty,0)$ in the $x-y$ plane,
otherwise the singularities are the lines $(x,\pm\infty)$ and $(\pm\infty,y)$
as in the C-metric. So the singularities for the rotating C-metric are
spinning rings (one is inside a black hole) since the singular points in the
$x-y$ plane are outside the axis and by the axial symmetry they must be rings.
Recall that for the C-metric the singularities are pieces of the axis (some
are inside the black holes).

The roots $w_{i}$ of $F(y)$ are the Killing horizons. The timelike Killing
vector field which is normal to the horizon is the linear combination
$\chi=\partial_{t}+\Omega_{H}\partial_{\phi}$ where the\ ``angular velocity''
of the horizon is
\[
\Omega_{H}=\left.  \Omega\right|  _{y=w_{i}}=-aA\left(  Kw_{i}\right)  ^{2}.
\]
and $\Omega\equiv-g_{tt}/g_{\phi t}$. The norm of this Killing vector (see
also \cite{Bicak99})
\begin{equation}
\chi^{\mu}\chi_{\mu}=\left(  \frac{K}{A\left(  x+y\right)  }\right)
^{2}\left\{  \left[  1+\left(  aAw_{i}x\right)  ^{2}\right]  ^{2}F(y)-\left[
aA\left(  w_{i}^{2}-y^{2}\right)  \right]  ^{2}G(x)\right\}  /\left\{
W\right\}
\end{equation}
vanishes at $y=w_{i}$. Note the rigid rotation of the black hole from the fact
the $\Omega_{H}$ is constant on the horizon. We can also compute the ``surface
gravity''. It has the same expression as in (\ref{surfgravb}). The area of
each horizon and the mass of the hole are also similar to the C-metric case
given by the equations (\ref{area}) and (\ref{mass}). Note however that the
roots $w_{1},w_{2}$ and $w_{3}$ of the quartic (\ref{Q4}) depend on the factor
$aA$. Although the expression of some quantities of the rotating C-metric are
similar the C-metric, they are not equal.

The norm of the Killing vector $\xi=A\partial_{t}$ is
\begin{equation}
\xi^{2}=\frac{F(y)-\left(  aA\right)  ^{2}G(x){y}^{4}}{\left[  KA\left(
x+y\right)  \right]  ^{2}W}%
\end{equation}
The roots of $\xi^{2}$ represent the boundaries of the surfaces of infinite
red-shift. The regions between the surfaces of infinite red-shift and the
Killing horizons are the ergoregions. The rotating regions are given by
\cite{Bicak99}
\[
F(y)G(x)\geq0.
\]
As in the case of the C-metric, we restrict to the cases of \ signature $-2$,
i.e. $G(x)\geq0$.

The Killing vector field
\[
\eta=\partial_{\phi}-\Omega_{H}\partial_{t}%
\]
has norm given by
\begin{equation}
\eta^{2}=\left(  \frac{1}{KA\left(  x+y\right)  }\right)  ^{2}\left\{
G(x)\left(  1+\left(  aAw_{i}y\right)  ^{2}\right)  ^{2}-\left[  aA\left(
x^{2}-w_{i}^{2}\right)  \right]  ^{2}F(y)\right\}  /\left\{  W\right\}
\end{equation}
One can prove, by polynomial analysis, that $\eta$ is a spacelike Killing
vector wherever $G(x)>0$ and $F(y)>0$. The axis of symmetry are given by
$x=-w_{i}$ where $\eta^{2}=0$, i.e. $G(-w_{i})=0$.

As in section 2, one can compute the ratio between the length of a circle by
$2\pi$ times its radius. If this ratio is not unity, there is an angle
depletion, that is, a conical singularity. Note however the dragging of the
inertial frame in virtue of the orbits of the spacelike Killing vector
$\eta=\partial_{\phi}-\Omega_{H}\partial_{t}$, that is, $K^{2}dt=aAw^{2}d\phi
$. Thus one has to compute the ratio from the metric (\ref{cmetricstat}).:
\begin{equation}
\lim_{\varepsilon\rightarrow0}\frac{\int_{0}^{2\pi}\frac{1}{A\left|
-w_{i}+\varepsilon+y\right|  }\frac{\sqrt{\left|  G(-w_{i}+\varepsilon
)\right|  }}{K}\left(  1+\left(  aAy\left(  -w_{i}+\varepsilon\right)
\right)  ^{2}\right)  d\phi}{2\pi\int_{-w_{i}}^{-w_{i}+\varepsilon}\frac
{1}{A\left|  x+y\right|  }\sqrt{\frac{1+\left(  aAyx\right)  ^{2}}{\left|
G(x)\right|  }}dx}=\frac{G_{x}\left(  -w_{i}\right)  }{2K}%
\end{equation}
One can choose the constant $K$ in such a way to avoid the conical singularity
in a particular piece of the axis. But in general the conical singularity will
show up somewhere on the axis. This is a manifestation of a spinning string singularity.

The angular velocity of the string at the roots $x=-w_{j}$ is given by
\begin{equation}
\Omega_{string}=\frac{K^{2}}{aw_{j}^{2}}%
\end{equation}
Thus in general, the string and the black holes have angular velocities with
different values and opposite senses.

Therefore, the picture of a piece of the $x-y$ plane with their interpretation
is shown in Figure 2.

The relative value of the invariant (\ref{C2b}) is shown in Figure 3. Note its
growing values as the singularity is approached.

One last remark. The total mass of the hole as given by
\begin{equation}
\kappa_{(i)}\mathcal{A}_{(i)}^{[k+1,k]}=4\pi\frac{m\left(  w_{k+1}%
-w_{k}\right)  }{2}=4\pi\;Mass
\end{equation}
now carries information on both acceleration and rotation since the roots
depend on those parameters.

Other versions of this solution \cite{FarhooshZimmerman80b} have similar features.

See the Appendix for the connection between the rotating C metric and the
Lewis-Papapetrou coordinates for vacuum stationary axisymmetric spacetimes.

\section{Discussions}

The C metric can represent several spacetimes depending on the range of the
coordinates. As shown in the table I, the spacetimes have singularities, event
horizons and conical singularities along the axis. If one takes appropriate
combinations of the rectangles in the table I, one gets one of the
interpretations found in the literature by some coordinate transformation.

By studying the geometrical quantities of the C metric we find the correct
interpretation for the spacetime it generates independently of a particular
transformation of coordinates.

This know-how can be of valuable help in the study of black hole accelerated
during a finite time. Interesting effects like dragging of inertial frame and
gravitational radiation are present.

The main conclusions of our study are: In general, the C metric and the
rotating C metric represent accelerated black holes with non smooth compact
horizon -- there is the possibility of toroidal-like black holes. It requires
a fine tuning of the constants to get a smooth compact horizon. In general the
axis of symmetry is not elementary flat. The surface gravity at the holes are
stronger than the frame acceleration. Therefore the accelerated black hole
temperature is higher than the temperature of the thermal bath associated to
the accelerated frame. The mass of the black hole can be computed from the
mechanics of black holes and it is the asymptotic mass of the Weyl solution of
a rod with line density $1/2$. We found no mathematical limitation on the
acceleration parameter. For the rotating case the mass has the contribution of
the rotation and the acceleration, as it should.

Although the solutions have some bizarre features, they give us lots of
informations on how the spacetime is dragged along an accelerated black hole.
The extension of uniqueness theorems to include accelerating black holes is
investigated in \cite{Wells98}

\bigskip

\textbf{Acknowledgments}

\bigskip

We want to thank FAPESP for financial support, also PSL is grateful to CNPq
for a research grant and to G. Gibbons for some early conversations about
different aspects for the C-metric.

\section{Appendix}

\subsection{C metric and Weyl coordinates}

One can improve our interpretation of the C metric through the transformation
from the coordinate $(t,x,y,\phi)$ into static axisymmetric spacetime in Weyl
like dimensionless coordinates $(t,r,z,\phi)$. This transformation is valid
only on the static regions of the $x-y$ plane (the even labeled regions of
Table I). Let the Weyl metric be
\begin{equation}
ds^{2}=m^{2}\left[  \exp\left(  2\psi\right)  dt^{2}-\exp\left(  2\nu
-2\psi\right)  \left[  dr^{2}+dz^{2}\right]  -r^{2}\exp\left(  -2\psi\right)
d\phi^{2}\right]  \label{weyl}%
\end{equation}
$m$ is the dimensional constant which settles the physical scale. The
functions $\psi$ and $\nu$ depend only on $r$ and $z$. From (\ref{weyl}) and
(\ref{cmetric}) one finds
\begin{align}
\left.
\begin{array}
[c]{c}%
\exp\left(  2\psi\right)  =\frac{K^{2}F(y)}{\left(  mA\right)  ^{2}\left(
x+y\right)  ^{2}}\\
r^{2}\exp\left(  -2\psi\right)  =\frac{G(x)}{\left(  KmA\right)  ^{2}\left(
x+y\right)  ^{2}}%
\end{array}
\right\}   &  \Rightarrow r^{2}=\frac{F(y)G(x)}{\left[  mA\left(  x+y\right)
\right]  ^{4}}\label{coordtransf}\\
\exp\left(  2\nu\right)  \left[  dr^{2}+dz^{2}\right]   &  =\frac{K^{2}%
F(y)}{\left(  mA\right)  ^{2}\left(  x+y\right)  ^{4}}\left[  \frac{dy^{2}%
}{F(y)}+\frac{dx^{2}}{G(x)}\right] \nonumber
\end{align}

One sees that the roots of $F(y)$ are linked to the regions, in Weyl
coordinates, where $\psi\rightarrow-\infty$ . Recall that the Einstein vacuum
equations for the Weyl metric (\ref{weyl}) reduce to
\begin{align*}
\nabla^{2}\psi &  \equiv\psi_{rr}+\frac{1}{r}\psi_{r}+\psi_{zz}=0\\
d\nu &  =r\left(  \psi_{r}^{2}-\psi_{z}^{2}\right)  \;dr+2r\psi_{r}\psi
_{z}\;dz
\end{align*}
Thus the function $\psi$ must be a solution of Laplace's equation arising from
sources lying on the axis $r=0$ \cite{CornishUttley95} and asymptotically
behaves as the Newtonian gravitational potential of those sources. Neglecting
the negative mass density cases one can show that the Newtonian sources for
the even labeled regions of the C-metric have mass density given by
\cite{CornishUttley95}
\[
\lim_{r\rightarrow0}\left(  \frac{1}{2}r\psi_{r}\right)  =\lim_{y\rightarrow
w_{i};dx=0}\frac{1}{2}\left(  \frac{F_{y}/F-2/(x+y)}{F_{y}/F-4/(x+y)}\right)
=\frac{1}{2}%
\]
It is known that semi-infinite line source and finite line source with mass
density $\frac{1}{2}$ are associated, through the Weyl solutions, to Rindler
and Schwarzschild spacetimes respectively \cite{Bondi57} \cite{Bonnor83}.
Thus, if the cubic (\ref{Q3}) has three distinct real roots, the roots of $F$
will be associated to the line sources and the roots of $G$ will be associated
to pieces of the $z$-axis. We can assign the points $z_{i}$ along the $z$-axis
in Weyl coordinates where the line sources begin or end in such a way that
$z_{i}=w_{i}$ so that $z_{i}$ are also the roots of the cubic $Q$.

The conical singularity in Weyl coordinate appears where
\begin{equation}
\lim_{\varepsilon\rightarrow0}\frac{\int_{0}^{2\pi}r\exp\left(  -\psi\right)
d\phi}{2\pi\int_{0}^{\varepsilon}\exp\left(  \nu-\psi\right)  dr}=1/\exp
\nu(0,z) \label{con2}%
\end{equation}
is not unity and the compression force on it is given
by\cite{LetelierOliveira98}
\begin{equation}
F_{z}=\frac{1}{4}\left[  \exp\left(  -\nu(0,z)\right)  -1\right]  =\frac{1}%
{4}\left[  \frac{G_{x}\left(  -w_{i}\right)  }{2K}-1\right]  \label{force}%
\end{equation}
where the second equality is obtained by the comparison of (\ref{con1}) and
(\ref{con2})

\subsection{Rotating C metric and Lewis-Papapetrou coordinates}

One can improve our interpretation of the rotating C metric by the comparison
of the coordinate system $(t,x,y,\phi)$ with the stationary axisymmetric
spacetime in Lewis-Papapetrou coordinates $(t,r,z,\phi)$. This comparison
holds only on the stationary regions. Let the metric be
\begin{equation}
ds^{2}=m^{2}\left[  \exp\left(  2\psi\right)  \left(  dt-\varpi d\phi\right)
^{2}-\exp\left(  2\nu-2\psi\right)  \left[  dr^{2}+dz^{2}\right]  -r^{2}%
\exp\left(  -2\psi\right)  d\phi^{2}\right]  \label{papa}%
\end{equation}
The functions $\psi$ and $\nu$ and $\varpi$ depend on $r$ and $z$ only and all
quantities but $m$ are dimensionless. From (\ref{papa}) and (\ref{cmetricstat}%
) one finds
\begin{align}
&  \left.
\begin{array}
[c]{c}%
\exp\left(  2\psi\right)  =\frac{K^{2}W^{-1}}{\left(  mA\right)  ^{2}\left(
x+y\right)  ^{2}}\left[  F(y)-\left(  aAy^{2}\right)  ^{2}G(x)\right] \\
r^{2}\exp\left(  -2\psi\right)  +\varpi^{2}\exp\left(  2\psi\right)
=\frac{G(x)-\left(  aAy^{2}\right)  ^{2}F(y)}{\left(  KmA\right)  ^{2}\left(
x+y\right)  ^{2}W}\\
\varpi\exp\left(  2\psi\right)  =\frac{aW^{-1}}{Am^{2}\left(  x+y\right)
^{2}}\left[  x^{2}F(y)+y^{2}G(x)\right]
\end{array}
\right\} \\
r^{2}  &  =\frac{F(y)G(x)}{\left[  mA\left(  x+y\right)  \right]  ^{4}%
}-\left(  \frac{a}{m}\right)  ^{2}\frac{\left[  1+\left(  mA\right)
^{2}\right]  \left[  x^{2}F(y)+y^{2}G(x)\right]  ^{2}}{W^{2}\left[  mA\left(
x+y\right)  \right]  ^{4}}%
\end{align}

One sees that the roots of $F(y)-\left(  aAy^{2}\right)  ^{2}G(x)$, the
infinite red-shift surfaces, are linked to the regions where $\psi
\rightarrow-\infty$ . The full transformation is very complicate and not
clarifying. The Einstein vacuum equations for the metric (\ref{papa}) can be
written as
\begin{align*}
\nabla^{2}\psi &  \equiv\psi_{rr}+\frac{1}{r}\psi_{r}+\psi_{zz}=-\frac
{\exp\left(  4\psi\right)  }{2r^{2}}\left(  \nabla\varpi\right)  ^{2}\\
\nu_{r}  &  =r\left(  \psi_{r}^{2}-\psi_{z}^{2}+\frac{\exp\left(
4\psi\right)  }{4r^{2}}\left(  \varpi_{z}^{2}-\varpi_{r}^{2}\right)  \right)
\\
\nu_{z}  &  =2r\left(  \psi_{r}\psi_{z}-\frac{\exp\left(  4\psi\right)
}{4r^{2}}\varpi_{z}\varpi_{r}\right) \\
0  &  =\nabla\cdot\left(  \frac{\exp\left(  4\psi\right)  }{r}\nabla
\varpi\right)
\end{align*}
where $\nabla$ stands for the flat vector operator $\left(  \partial
_{r},\partial_{z}\right)  $. Thus the function $\psi$ must be a solution of
the non-linear Poisson's equation which have as the source a contribution from
the rotation potential $\varpi$. The connection between the solutions of the
equations above and the rotating C-metric solution is not simple. Nevertheless
it is known that there are soliton solutions associated to Newtonian images of
semi-infinite line plus a finite line with mass density $\frac{1}{2}$ that
represent the rotating version of the Weyl C-metric \cite{Letelier85}.

\newpage

\begin{center}%
\begin{figure}
[pb]
\includegraphics[
height=10.3132cm,
width=10.1067cm
]%
{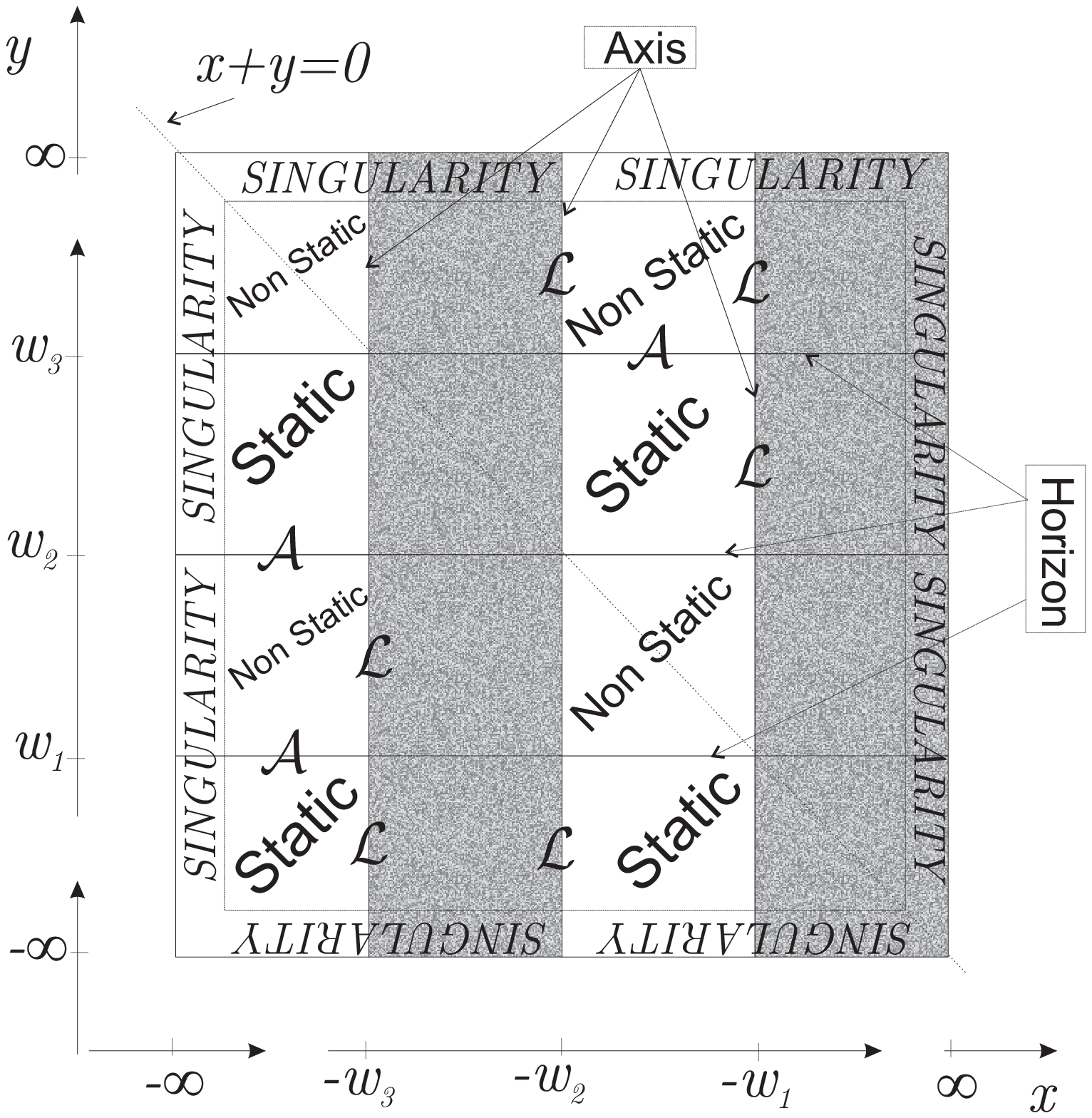}%
\caption{The $x-y$ plane for the C-metric. The vertical separator lines are
axis and the horizontal ones are horizons. For comparison see similar figure
in \protect\cite{CornishUttley95}.
}
\end{figure}
\end{center}

\pagebreak %

\begin{figure}
[ptb]
\begin{center}
\includegraphics[
height=17.1886cm,
width=13.3247cm
]%
{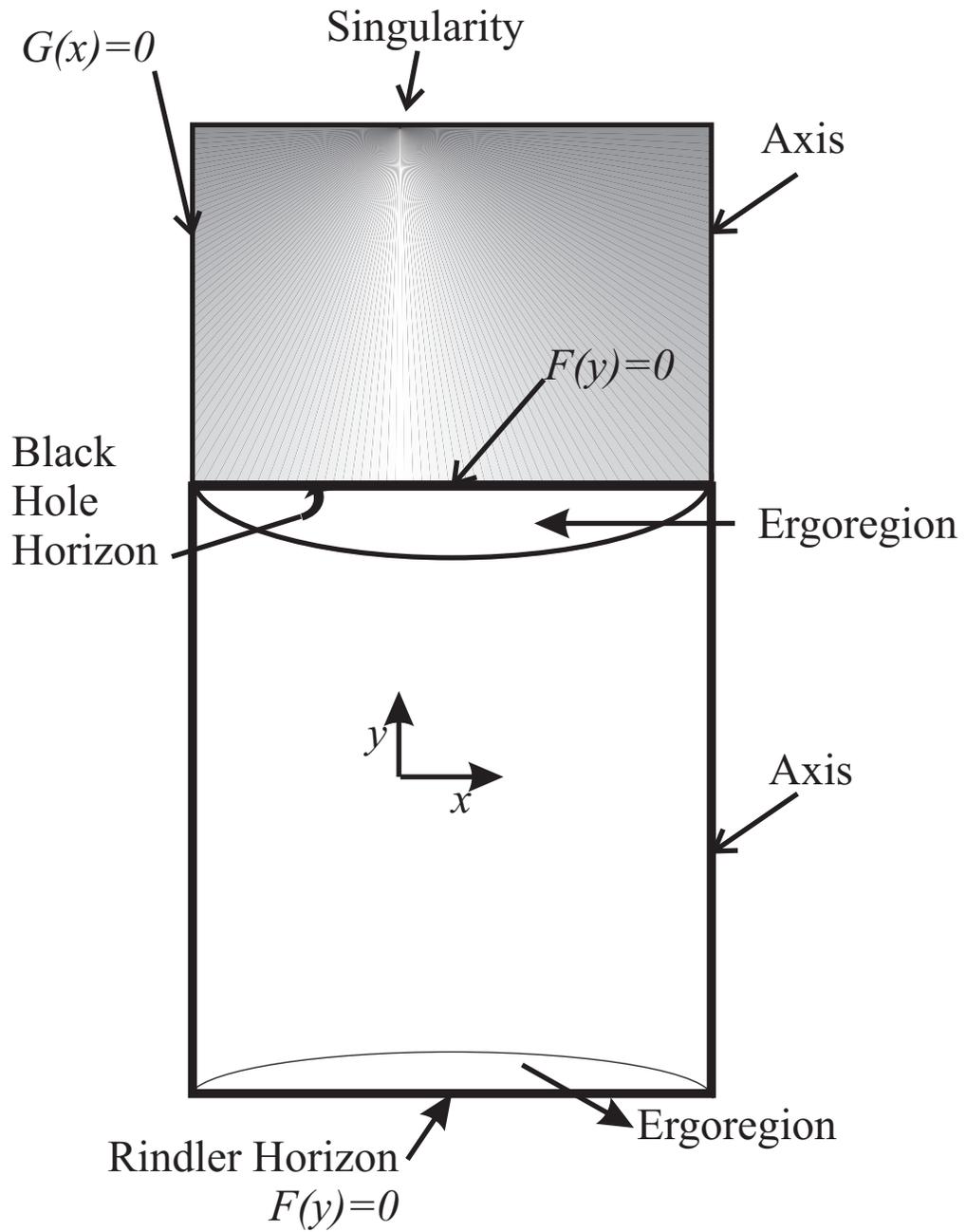}%
\caption{A piece of the $x-y$ plane for the Stationary C-metric. The axis,
horizons and the ergo-regions are displayed.}%
\end{center}
\end{figure}

\pagebreak %

\begin{figure}
[ptb]
\begin{center}
\includegraphics[
height=12.5186cm,
width=11.3192cm
]%
{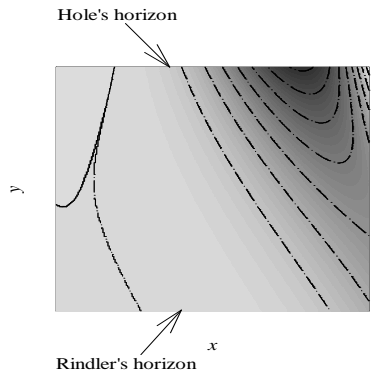}%
\caption{Contour plot of the relative values of the invariant (\ref{C2b}) for
an outter domain of comunication of the Rotating C-metric (\ref{cmetricstat}%
). The parameters are the following: $w_{1}=-1$, $w_{2}=1$, $w_{3}=2$, $m=1$,
$K=1$, $a=1/2$ and $A=1/3$.}%
\end{center}
\end{figure}

\end{document}